\documentstyle[a4,12pt]{article}

\newcommand{\be}{\begin{equation}}
\newcommand{\ee}{  \end{equation}}
\newcommand{\bea}{\begin{eqnarray}}
\newcommand{\eea}{  \end{eqnarray}}

\newcommand{\Author}[1]{#1,}
\newcommand{\Title}[1]{{\em #1},}
\newcommand{\Journal}[1]{#1}

\newcommand\balpha{{\mbox{\protect\boldmath$\alpha$}}}
\newcommand\balphas{{\mbox{\scriptsize\protect\boldmath$\alpha$}}}
\newcommand\balphah{{\mbox{\protect\boldmath$\hat\alpha$}}}
\newcommand\bbeta{{\mbox{\protect\boldmath$\beta$}}}
\newcommand\bbetas{{\mbox{\scriptsize\protect\boldmath$\beta$}}}

\newcommand\ba{{\mbox{\protect\boldmath$a$}}}
\newcommand\bas{{\mbox{\scriptsize\protect\boldmath$a$}}}
\newcommand\bab{{\mbox{\protect\boldmath$\bar a$}}}
\newcommand\babs{{\mbox{\protect\scriptsize\boldmath$\bar a$}}}
\newcommand\bb{{\mbox{\protect\boldmath$b$}}}
\newcommand\bbs{{\mbox{\scriptsize\protect\boldmath$b$}}}

\newcommand\bB{{\mbox{\protect\boldmath$B$}}}
\newcommand\bBs{{\mbox{\scriptsize\protect\boldmath$B$}}}
\newcommand\bBb{{\mbox{\protect\boldmath$\bar B$}}}
\newcommand\bBbs{{\mbox{\protect\scriptsize\boldmath$\bar B$}}}
\newcommand\bC{{\mbox{\protect\boldmath$C$}}}
\newcommand\bc{{\mbox{\protect\boldmath$c$}}}
\newcommand\bcs{{\mbox{\scriptsize\protect\boldmath$c$}}}

\newcommand\bo{{\mbox{\protect\boldmath$0$}}}
\newcommand\bu{{\mbox{\protect\boldmath$u$}}}
\newcommand\bub{{\mbox{\protect\boldmath$\bar u$}}}
\newcommand\bv{{\mbox{\protect\boldmath$v$}}}
\newcommand\bw{{\mbox{\protect\boldmath$w$}}}
\newcommand\bphi{{\mbox{\protect\boldmath$\phi$}}}
\newcommand\bTb{{\mbox{\protect\boldmath$\bar T$}}}

\begin{document}
\begin{titlepage}
\hfill{hep-th/9404072}

\vspace*{\fill}
\centerline{\huge On the form of local conservation laws}
\centerline{\ }
\centerline{\huge for some relativistic field theories}
\centerline{\ }
\centerline{\huge in 1+1 dimensions.}

\vspace{2cm}
\centerline{\large
Erling G. B. Hohler and K\aa re Olaussen}

\vspace{.2cm}
\centerline{\large Institutt for fysikk, NTH,}
\vspace{.1cm}
\centerline{\large Universitetet i Trondheim}
\vspace{.1cm}
\centerline{\large N--7034 Trondheim, Norway.}

\vspace{.5 cm}
\centerline{April 13, 1994}

\vspace{.5cm}
\begin{abstract}
We investigate the possible form of local translation invariant
conservation laws associated with the
relativistic field equations $\partial\bar\partial\phi_i=-v_i(\bphi)$ for a
multicomponent field $\bphi$. Under the assumptions that
(i)~the $v_i$'s
   can be expressed as linear combinations of partial derivatives
   $\partial w_j/\partial\phi_k$ of a set of functions $w_j(\bphi)$,
(ii)~the space of functions spanned by the $w_j$'s is closed under
   partial derivations,  and
(iii)~the fields $\bphi$ take values in a simply connected space,
the local conservation laws can either be transformed to the form
$\partial{\bar{\cal P}}=\bar\partial\sum_j w_j {\cal Q}_j$ (where
$\bar{\cal P}$ and ${\cal Q}_j$ are homogeneous polynomials in the variables
$\bar\partial\phi_i$, $\bar\partial^2\phi_i$,\ldots), or to the parity
transformed version
of this
expression
$\partial\equiv(\partial_t+\partial_x)/\sqrt{2}\rightleftharpoons\bar\partial
\equiv (\partial_t-\partial_x)/\sqrt{2}$.
\end{abstract}

\vspace*{\fill}

\end{titlepage}

\section{Introduction}

As has been known at least since the days of Liouville\cite{Liouville}
there exist exactly solvable non-linear field theories in
1+1 space-time dimensions. During the last few decades the knowledge
of such models has increased dramatically. Examples of current
interest are the Toda and affine Toda field
theories\cite{LezSav,Mansfield}. These are (multi-component)
field models with relativistic (and additional) symmetries. There
is essentially one such model for each Lie algebra\cite{Humpreys}
or affine Lie algebra\cite{Kac}.
The ordinary Toda field theories are conformally invariant, while
each affine Toda field theory may be viewed as a perturbation of
an ordinary model away from its conformal limit.
The simplest examples
of such models are respectively the Liouville equation\cite{Liouville},
\be
  \partial\bar\partial\phi = -\mbox{e}^\phi,
  \label{Liouville}
\ee
and the sinh-Gordon equation,
\be
  \partial\bar\partial\phi=-\mbox{e}^\phi+\varepsilon\,\mbox{e}^{-\phi},
  \label{sineGordon}
\ee
which formally can be viewed as a perturbation of the Liouville equation.
(After the transformations $\phi \to \phi+\frac{1}{2}\log\varepsilon$,
$\left(\partial,{\bar\partial}\right) \to 2\sqrt{\varepsilon}\left(\partial,
{\bar\partial}\right)$ equation (\ref{sineGordon}) acquires the conventional
form $\partial\bar\partial \phi = - \sinh\phi$.)

Since two-dimensional statistical models (with local interactions)
can be described by conformal field theories at their
critical points\cite{BPZ,Cardy}, the (affine) Toda
field theories may be useful for modelling the relevant degrees of
freedom of many $2d$ statistical
models at, and in the vicinity of, their critical points. It has been shown
that the Toda field theories can be chosen to have critical
exponents in agreement with the known rational exponents of a large class of
such models\cite{HolloMans,Fendley et al}. This, of course, requires
quantization or functional integral formulation, followed by a Wick rotation
to Euclidean space. The (affine) Toda field theories also provide
interesting examples of interacting relativistic quantum field theories.

Consider now the classical theories.
As an integrable system each (affine) Toda field theory
has an infinite set of independent conserved quantities in involution
(so that the classical Poisson bracket of any two members of the set
vanishes). Since the field theory is local, one expects the conserved
quantities to be expressible as the integrated charges
\[
   Q = \int dx\,\left(X-{\bar X}\right)
\]
of a set of local conservation laws,
\be
    \bar\partial X = \partial {\bar X}.
    \label{conservationlaw}
\ee
Olive and Turok\cite{Olive et Turok} have found a method
for calculating an infinite set of local conservation
laws for most of the
affine Toda field theories, and have shown that the corresponding
charges are in involution. It is a more subtle question to decide
whether the set generated by this method is complete. Provided it
is, it might be used to construct a complete
set of action variables for the model.
(For a complete description the canonically
conjugate angle variables would also have to be found.)

It is believed that the integrability of the Toda field theories
in some sense survives quantization\cite{Sasaki} (although it in
general is rather unclear what should be meant by
`quantum integrability'\cite{Weigert}). In the
process of quantizing these models the problems of
renormalization and operator ordering are encountered.
In order to investigate these problems
we felt it would be useful to have explicit expressions for some
conserved currents beyond the energy-momentum tensor. For this
purpose we first attempted to use the method of Olive and
Turok\cite{Olive et Turok} for explicit computations. However, we
found it {\em very} cumbersome to proceed beyond the energy-momentum
tensor.
(This may perhaps be entirely due to our ineptness, but somewhat
to our comfort there does not seem to be anyone else in the literature who
has succeded either.)
In the end we found it simpler to use a rather pedestrian
direct method, where we by systematic algebraic reduction
narrow the possible forms of the conservations
laws (of a given class) until a (possibly
empty) set of solutions is found.
The prospects for proving the existence of infinitely many
conservation laws with such a procedure are of course
rather meagre,
but the method works quite satisfactory for the first few conservation
laws---in particular with the assistance of an algebraic manipulation
program.
The procedure should be viewed as supplementary to the powerful method
in reference \cite{Olive et Turok}, whose results also provides
great encouragement for the searching process---since the sought-for
quantities are already known to exist.
But our method can also be applied to cases which are not
covered by the method \cite{Olive et Turok}, and we have indeed discovered
additional conservation laws through it.

The purpose of this paper is to perform the first step of the
reduction process mentioned above, for a quite general class
of models. This reduction is based on the fact that on the
classical level there are many essentially equivalent ways of writing a
conservation law (\ref{conservationlaw}), by modifications of the form
\be
   X \to X + \partial Y,\ \ \
  {\bar X} \to {\bar X} + {\bar\partial} Y.
  \label{redef}
\ee
We shall refer to this process as the addition
of trivial conservation laws (although some of these `trivial'
laws in some  Toda field theories may integrate to a non-zero
topological charge).

Using this freedom alone we are able to restrict the form of the
conservation laws to the extent stated in the Abstract. The
remainder of this paper is devoted the proof of these results.
The next section is mainly of illustrative purpose; here we first
apply our procedure to the Liouville model, and next show how the
obtained results can be generalized to a much wider class
of one-component field theories. Following this discussion,
the general case of multi-component fields is
treated in the section 4.

We have used the results of this paper as the starting point for
finding explicit expressions for the conservation laws
in the ordinary and affine Toda field theories \cite{HO1,HO2}.

\section{Example: The Liouville Model}

We shall in this section introduce our notation and basic ideas
in the simplest example of a Toda field theory, i.e.\ the
Liouville model---defined by a one-component field $\phi$ satisfying the
equation of motion (\ref{Liouville}),
where $\partial\equiv 2^{-1/2}(\frac{\partial}{\partial t}+
\frac{\partial}{\partial x})$ and
${\bar\partial}\equiv
2^{-1/2}(\frac{\partial}{\partial t}-\frac{\partial}{\partial x})$.
{\em A priori},
a local translation invariant conservation law for this model must be of
the form (\ref{conservationlaw}),
where $X$ and $\bar X$ depend on $(t,x)$ through
the variables $\bu \equiv (\partial\phi,\partial^2\phi,
\partial^3\phi,\dots)$, $\bub \equiv
({\bar\partial}\phi,{\bar\partial}^2\phi,
{\bar\partial}^3\phi,\dots)$, and $v$.
Variables involving mixed
derivatives, $\partial^m {\bar\partial}^n\phi$ with $m n\ne 0$, can
be eliminated by (repeated) use of the equation of motion (\ref{Liouville}).
We consider the case that $X$ and ${\bar X}$ are finite
polynomials in $\bu$, $\bub$, and $v$. Since
equation (\ref{Liouville}) is invariant under the rescalings,
\be
  \partial \to  \Lambda\,\partial,\ \ \
  {\bar\partial} \to {\bar\Lambda}\,{\bar\partial},\ \ \
  v \to \Lambda{\bar\Lambda}\,v,
\ee
we may restrict our attention to currents which transform irreducibly under
these, i.e.\ pairs $(X,{\bar X})$ such that
\be
  X \to \Lambda^{h}{\bar\Lambda}^{{\bar h}-1}\,X,\ \ \
  {\bar X} \to \Lambda^{h-1}{\bar\Lambda}^{\bar h}\,
  {\bar X}.
\ee
Thus $X$ resp.\ $\bar X$ must be a homogeneous polynomial of order
$(h,{\bar h}-1)$ resp.\ $(h-1,{\bar h})$, counted according to
the rules that each $u_j=\partial^j\phi$ carries
(conformal) weight $(j,0)$, each ${\bar u}_j$ weight $(0,j)$, and each
$v$ weight $(1,1)$.

Consider now a conservation law for which $h \le {\bar h}$ (the case
${\bar h} < h$ follows by interchanging $\partial$ and
$\bar\partial$ throughout). We shall show that by appropriate
(and possibly repeated) modifications (\ref{redef})
all dependency on the variables ${\bf u}$ may be removed from the
expressions for the currents, and all dependency on $v$ may be removed
from the expression for $\bar X$.

Focussing attention on the ${\bf u}$ dependence we may write
\be
     {\bar X} = \sum_{\balphas}
     {\bar C}_{\balphas}(\bub,v)\,
     u_h^{\alpha_{h-1}}\cdots u_2^{\alpha_2}\,u_1^{\alpha_1} \equiv
     \sum_{\balphas} {\bar C}_{\balphas}\,{\bu}^{\balphas},
     \label{Xpol}
\ee
where the sum runs over vectors
$\balpha=(\alpha_{h-1},\ldots,\alpha_2,\alpha_1)$
of integers $\alpha_j\ge0$ such that $\sum_{j=1}^h j\alpha_j \le {h-1}$.
There is a similar expression for $X$.
We can arrange the vectors  in
lexical order (which is a total ordering),
and identify the term in $\bar X$ with the largest
$\balpha$-vector,
\be
  {\balphah}({\bar X})\equiv{\bab}.
\ee
We generally let ${\balphah}(\cdot)$ denote the functional
which extracts the largest
$\balpha$-vector from a given expression.
The term referred to above is ${\bar C}_{\babs} {\bu}^{\babs}$.
Likewise, we can identify the term in $X$ with the largest $\balpha$-vector,
\be
  {\balphah}(X)\equiv \ba,
\ee
i.e.\ the term $C_{\bas} {\bu}^{\bas}$.

Note that if
$\balpha =(0,\ldots,0,\alpha_j,\ldots,\alpha_1) > (0,\ldots,0,0) \equiv \bo$,
with $j\le h$ the largest integer for which $\alpha_j >0$
($j\le h-1$ in the case of $\bar X$), then
\bea
   &&\partial\balpha \equiv {\balphah}({\partial\bu}^{\balphas}) =
   (0,\ldots,1,\alpha_j-1,\dots,\alpha_1) > \balpha,\\
   &&{\bar\partial}\balpha \equiv {\balphah}({\bar\partial \bu}^{\balphas})
   = (0,\ldots,0,\alpha_j,\ldots,\alpha_1-1) < \balpha,
\eea
(with rather obvious modifications of the explicit expressions
if $j=1$ or $\alpha_1=0$). For now, assume that $\bab > \bo$. Then,
the term in $\partial{\bar X}$ with the largest
$\balpha$-vector is contained in the expression
\(
  {\bar C}_{{\babs}}\,\partial {\bu}^{\babs}
\),
and the term in ${\bar\partial}X$ with the largest $\balpha$-vector
is contained in the expression
\(
   \left({\bar\partial} C_{\bas}\right){\bu}^{\bas}
\).
Note that we are assured to have ${\bar\partial} C_{\bas} \ne 0$, since by
assumption
${\bar h} \ge h > 0$, the last inequality being a consequence of $\bab > \bo$.
Since $\partial{\bar X}={\bar\partial}X$, we must have that
\be
   \partial{\bab}= \ba.
\ee

The statements above can be made more concise by introducing
a notation for calculations which are
exact only to leading order in the $\balpha$-sequence:
Let
\be
    A \equiv \sum_{\balphas} C_{\balphas}(A) {\bu}^{\balphas} \simeq
    B \equiv \sum_{\balphas} C_{\balphas}(B) {\bu}^{\balphas}
\ee
denote that ${\balphah}(A) = {\balphah(B)} \equiv \bc$, and that
$C_{\bcs}(A) = C_{\bcs}(B)$.
Thus, we have that
\be
    X \simeq C_{\bas} {\bu}^{\bas},\ \ \ \
    {\bar X} \simeq {\bar C}_{\babs} {\bu}^{\babs},
\ee
and, for $\bab > \bo$,
\bea
    &&\partial{\bar X} \simeq {\bar C}_{\babs} \partial{\bu}^{\babs},
    \nonumber\\&&
    {\bar\partial}X \simeq
    \left({\bar\partial} C_{\bas}\right){\bu}^{\bas} \simeq
    \left({\bar\partial}C_{\bas}\right)\partial\bu^{\babs} \simeq
    {\bar\partial}\left( C_{\bas} \partial {\bu}^{\babs}\right) \simeq
    {\bar\partial}\partial\left( C_{\bas} {\bu}^{\babs}\right).
    \label{trivial}
\eea
The last two equivalences are true because of the inequalites
$\balphah({\bar\partial}\partial u^{\babs}) < \balphah(\partial u^{\babs})$,
and $\balphah(u^{\babs}) < \balphah(\partial u^{\babs})$.
Equation (\ref{trivial}) means that the leading order part of the conservation
law is
trivial, and may be removed by a redefinition (\ref{redef}).
This process may be continued until we end up with $\bab=\bo$ after a
finite number of steps.

Thus, we have shown that $\bar X$ can be reduced to a polynomial in
$\bub$ and $v$. It remains to show that also the
dependency on $v$ can be eliminated by the addition of trivial
conservation laws. It turns out that we may equally well assume the
more general form
\(
   {\bar X} = {\bar X}(\bub,\phi)
\),
and show that an arbitrary dependency on $\phi$ can be eliminated.
Assuming this form we find
\be
   \partial{\bar X} \simeq
   \left({\partial{\bar X}}/{\partial\phi}\right) \partial\phi,
\ee
provided ${\partial{\bar X}}/{\partial\phi}\ne 0$.
This should be identified with
\be
  {\bar\partial}X \simeq \left({\bar\partial} C_{\bas}\right)\partial\phi
  \simeq {\bar\partial}\partial D,
\ee
where
\[
    D(\bub,\phi) = \int^{\phi} d\phi'\, C_{\bas}(\bub,\phi')
\]
is a well-defined function provided $\phi$ takes values in a
simple-connected space.
Thus, the order $\partial\phi$-part of the conservation law above can be
eliminated by the addition of a trivial conservation law,
$X \to X - \partial D$, ${\bar X} \to {\bar X}-{\bar\partial}D$,
leaving us in
the promised situation where ${\bar X}={\bar{\cal P}}(\bub)$.
{}From this form it
follows by the use of the equation of motion (\ref{Liouville}) that
\[
  \partial{\bar X} = \sum_i
  \left(\partial{\bar X}/\partial{\bar u}_i \right)\partial{\bar u}_i =
  v {\bar{\cal Q}}(\bub),
\]
where $\bar{\cal Q}$ is a homogeneous polynomial of rank ${\bar h}-1$. This
in turn requires $X$ to be of the form
\be
   X = v {\cal Q}(\bub),
   \label{Xcurrent}
\ee
so that
\[
   {\bar\partial}X = v \left({\bar u}_1 +
   \sum_i {\bar u}_{i+1}\frac{\partial}{\partial{\bar u}_i}\right)
   {\cal Q}(\bub).
\]
Thus, the conservation law (\ref{conservationlaw}) is satisfied if and only
if
\be
  {\bar{\cal Q}}(\bub) = \left({\bar u}_1 +
  \sum_{i} {\bar u}_{i+1}\frac{\partial}{\partial{\bar u}_i}\right)
  {\cal Q}(\bub).
\ee
Before closing this section we stress that
the purpose of our presentation has been to
introduce and illustrate the methods to be used on the general case below,
not to find the already well known conservation laws for the Liouville model.
Since the (conformally improved) energy momentum tensor for the Liouville model
\[
  {\bar T} = \frac{1}{2}{\bar u}_1^2-{\bar u}_2 =
  \frac{1}{2}({\bar\partial}\phi)^2-{\bar\partial}^2\phi,
\]
satisfies the conservation law $\partial{\bar T}=0$,
it follows that any polynomial $\bar{\cal Q}(\bTb)$ in the variables
$\bTb \equiv ({\bar T},{\bar\partial}{\bar T},
{\bar\partial}^2{\bar T},\ldots)$ also satisfies
$\partial\bar{\cal Q}(\bTb) = 0$.

The reader should further note that almost no use have been made of the
explicit expression for the force term $v=v(\phi)$. We mainly
needed the fact that
$\partial v/\partial\phi \propto v$ to derive the explicit form
(\ref{Xcurrent}) for the current $X$.

Therefore, the results of this
section are
easily generalized to more general forces $v(\phi)$, i.e.\ those
which are the gradients, $v=-\partial w/\partial\phi$,
of a potential $w$ which satisfies an
homogeneous, finite order, constant coefficient
differential equation,
\[
  \left( \frac{\partial^{n+1}}{\partial\phi^{n+1}} + c_n
  \frac{\partial^{n}}{\partial\phi^{n}} + \cdots + c_0 \right) w(\phi)=0.
\]
This includes all polynomial potentials.
Then all translation invariant conservation laws (\ref{conservationlaw})
can be transformed to a form where
\be
    {\bar X} = {\bar{\cal P}}({\bar u}),\ \ \ \
    X = \sum_{j=0}^n w_j\,{\cal Q}_j({\bar u}),
\ee
with an $(n+1)$-component vector
\[
   \bw \equiv \left(w,\partial w/\partial\phi,
   \ldots,\partial^n w/\partial\phi^n\right).
\]
This, of course, agrees with the general form of the conserved
energy momentum tensor,
\[
  {\bar X} = \frac{1}{2}{\bar u}_1^2-
  \alpha\,{\bar u}_2,\ \ \ \
   X = w_0-2\alpha\, w_1,
\]
where $\alpha$ can
be adjusted arbitrarily by the addition of a trivial conservation law.
It is also in agreement with the infinitely many conservation laws for
a free field theory, $v(\phi) = \phi$, where e.g.\ the next order
conservation law reads
\be
  {\partial}{\bar u}_2^2 =
  -{\bar\partial} {\bar u}_1^2,
\ee
and sinh-Gordon
model, $v(\phi)=\sinh\phi$,
where e.g.\ the next order conservation law reads
\be
  {\partial}\left(\frac{1}{2}{\bar u}_1^2 - {\bar u}_2 \right)^2 =
  -{\bar\partial}\left(\mbox{e}^{-\phi}\,{\bar u}_1^2\right).
\ee
Note that these expressions exemplify the facts that
\begin{enumerate}

\item[(i)]
some of the conservation laws for the conformal theory can be obtained as a
limiting
case of the affine conservation laws, by shifting the $\phi$-field by an
infinite constant,

\item[(ii)]
some of the conservation laws for the free field theory can be obtained as a
limiting
case of the affine conservation laws, by extracting the pieces which are of
leading order as $\phi\to 0$,

\item[(iii)]
the conservation laws for the sine-Gordon model can be obtained from the
sinh-Gordon
conservation laws by making the replacement $\phi\to i\phi$. This will lead to
expressions
which involve complex currents, but where the real and imaginary parts must be
separately
conserved. However, in each case only one of the parts
constitutes a nontrivial conservation law.

\end{enumerate}

Since our procedure is of a restrictive rather than constructive nature,
the question whether there exist models beyond the examples above
with additional conservation laws
is left open.
But by investigating limiting behaviours like those above, the
possible form of such laws can be further
constrained in each specific model.

\section{The general case}

In this section we show that the results of the previous section can
be generalized to the models of a
$m$-component field $\bphi=(\phi_1,\ldots,\phi_m)$
satisfying the equation
\be
  \partial{\bar\partial}\bphi = - \bv = -{\bC}\,{\bw}
  \label{geneqn}
\ee
where $\bC$ is a $m\times n$ constant matrix, and
each component of $\bw=(w_1,\ldots,w_n)$ is an exponential expression
\be
  w_i = \exp{\textstyle\left(\sum_{j=1}^{m}
  k_{ij}\phi_{j} \right)}.
\ee
The most interesting special cases of (\ref{geneqn}) are the Toda
resp.\ the affine Toda field equations (one equation for each
Lie resp.\ affine Lie algebra). These are integrable systems with an
infinite number of conservation laws. However, equation
(\ref{geneqn}) is more than sufficient to restrict the form of
the conserved currents to the level we intend in this paper.
The analysis turns out to be a
modest extention of our example in the previous section---most of the
effort goes into establishing and explaining notation.

In a local translation invariant conservation law
${\bar\partial}X=\partial{\bar X}$ the currents $X$, $\bar X$ will
{\em a priori} depend
on $(x,t)$ through the matrices $\bu$ ($u_{ij}\equiv {\partial^j}\phi_i$),
$\bub$ (${\bar u}_{ij}\equiv{\bar\partial}^j\phi_i$), and the vector $\bw$.
Equation (\ref{geneqn}) is formally invariant under rescalings
\[
  \partial\to\Lambda\,\partial,\ \ \ \
  {\bar\partial}\to{\bar\Lambda}\,{\bar\partial},\ \ \ \
  \bw \to \Lambda\,{\bar\Lambda}\,\bw,
\]
although the latter is not necessarily implementable through a
transformation on the fields $\bphi$. However, a true symmetry
is not necessary for our purposes. We only need a power counting
assignment which is invariant under replacements like
\[
  \partial{\bar\partial}\bphi\to -\bv = -\bC\,\bw,\ \ \ \
  \partial w_i \to w_i \sum_j k_{ij}\,u_{j1},\ \ \ \
  {\bar\partial} w_i \to w_i \sum_j k_{ij}\,{\bar u}_{j1}.
\]
Thus we may assign $u_{ij}$
a weight $(j,0)$, ${\bar u}_{ij}$ a weight $(0,j)$, $w_i$ a
weight $(1,1)$, and consider each conserved current to be a homogenous
polynomial in these variables, $X$ of weight $(h,{\bar h}-1)$ and $\bar X$
of weight $(h-1,{\bar h})$. It is sufficient to consider the case
$h\le{\bar h}$, since the opposite case follows by interchanging
$\partial$ and $\bar\partial$ in all expressions.
We begin by demonstrating that all dependence on the
variable $\bu$ may be eliminated from ${\bar X}$ by the addition of trivial
conservation laws. Focussing on the $\bu$-dependence we write
\be
   {\bar X} = \sum_{\bbetas}\,{\bar C}_{\bbetas}(\bub,\bw)\,{\bu}^{\bbetas}
   \equiv
   \sum_{\bbetas}\,{\bar C}_{\bbetas}(\bub,\bw)\,
   \prod_{ij} u_{ij}^{\beta_{ij}},
\ee
where $\bbeta$ is a matrix of non-negative integers.
For each $\bbeta$ we define the vector
$\balpha(\bbeta)$ of integers by $\alpha_j =\sum_i \beta_{ij}$, and
conversely for each vector $\ba$ we define ${\cal B}(\ba)$ to be the inverse
image
of this mapping:
${\cal B}(\ba) \equiv \left\{ \bbeta \,\vert\, \balpha(\bbeta) = \ba \right\}$.
We again order the $\balpha$'s lexically, and let $\balphah(\cdot)$ denote the
functional which extracts the largest $\balpha$-vector from a given expression.
Thus, with
\be
   \ba = \balphah(X),\ \ \ \ \bB = {\cal B}(\ba),\ \ \ \
   \bab = \balphah({\bar X}),\ \ \ \ \bBb = {\cal B}(\bab),
\ee
we have that
\be
  X \simeq  \sum_{\bbetas \in \bBs}\, C_{\bbetas}\,{\bu}^{\bbetas},\ \ \ \
  {\bar X} \simeq \sum_{\bbetas \in \bBbs} {\bar C}_{\bbetas}\,{\bu}^{\bbetas}
\ee
to leading order in the {\balpha}'s.
By analysing the action of $\partial$ and $\bar\partial$ one realizes that
\be
   \balphah({\bar\partial}{\bu}^{\bbetas}) <
   \balphah({\bu}^{\bbetas}) <
   \balphah(\partial{\bu}^{\bbetas}),
   \label{balphaineq}
\ee
provided $\balphah(\bu^{\bbetas}) > \bo$.

For now, assume that ${\bab} > \bo$.
For each $\bbeta\in \bBb$ let
$\partial\bbeta$ be the set of $\bbeta'$-matrices such that $\partial
{\bu}^{\bbetas}$
contain a term proportional to ${\bu}^{\bbetas'}$, i.e.\ so that
\[
  \partial{\bu}^{\bbetas} = \sum_{\bbetas'\in\partial\bbetas}
  c_{\bbetas\bbetas'}\,{\bu}^{\bbetas'},
\]
with all the (integer) coefficients $c_{\bbetas\bbetas'} > 0$.
Further let
\be
   \partial\bb(\bbeta) =  \left\{ \bbeta'\in\partial\bbeta\,\vert\,
   \balpha(\bbeta') = \balphah(\partial\bbeta) \right\},
\ee
so that
\be
  \partial{\bu}^{\bbetas} \simeq \sum_{\bbetas'\in\partial\bbs(\bbetas)}
  c_{\bbetas\bbetas'}\,{\bu}^{\bbetas'},
 \label{derivative}
\ee
and let
\be
   \partial\bBb \equiv \bigcup_{\bbetas\in\bBbs} \partial\bb(\bbeta).
  \ee
It turns out that the sets $\partial\bb(\bbeta)$ are disjoint,
$\partial\bb(\bbeta_1) \cap \partial\bb(\bbeta_2) = \emptyset$ if
$\bbeta_1\ne\bbeta_2$,
so that for any element in $\bbeta' \in \partial\bBb$
one may reconstruct the $\bbeta$
from which it arose.
This can be seen as follows: Assume that $\bbeta$
is a $r\times n$ matrix, with at least one non-zero entry in the $n$'th column.
For each entry $\beta_{in}\ne 0$ of $\bbeta$, there exists a $r\times(n+1)$
matrix
$\bbeta'$ in $\partial\bb(\bbeta)$ with the properties that
\begin{enumerate}

\item[(i)]
$\bbeta'_{i,n+1}=1$,

\item[(ii)]
$\bbeta'_{in}=\bbeta_{in}-1$,

\item[(iii)]
$\bbeta'_{j,n+1}=0$ for all $j\ne i$,

\item[(iv)]
$\bbeta'_{jm} = \bbeta_{jm}$ for all $j\ne i$, $m\ne n$.

\end{enumerate}
\noindent
Thus, for any $\bbeta'$ in $\partial\bBb$ the corresponding $\bbeta$
is found by adding the last (non-vanishing) column to the next but last
column.

We then have that
\be
   {\partial}{\bar X} \simeq \sum_{\bbetas\in\bBbs}\, {\bar C}_{\bbetas}\,
   \partial\bu^{\bbetas}
   \simeq
   \sum_{\bbetas\in\bBbs\atop{\scriptstyle \bbetas'\in\partial\bbs(\bbetas)}}
\,
   {\bar C}_{\bbetas}\,c_{\bbetas\bbetas'} \bu^{\bbetas'} =
   \sum_{\bbetas'\in\partial\bBbs} {\tilde C}_{\bbetas'} {\bu}^{\bbetas'},
   \label{dXbar}
\ee
where ${\tilde C}_{\bbetas'}\equiv{\bar C}_{\bbetas}\, c_{\bbetas\bbetas'}$ (no
summation convention).
Likewise
\be
   {\bar\partial} X \simeq \sum_{\bbetas\in\bBs} \,
   \left({\bar\partial}{C}_{\bbetas}\right) \bu^{\bbetas}.
   \label{dbarX}
\ee
Thus, by comparing (\ref{dXbar}) and (\ref{dbarX}) one finds
that $\partial\bBb=\bB$, and
\be
  {\bar\partial}C_{\bbetas'} = {\tilde C}_{\bbetas'} =
  {\bar C}_{\bbetas} c_{\bbetas\bbetas'}\ \ \ \mbox{(no summation convention),}
  \label{coeffrelation}
\ee
since ${\bar\partial}{X} = {\partial}{\bar X}$ by assumption.

For each
$\bbeta'\in\partial\bb(\bbeta)$ let
${\hat C}_{\bbetas\bbetas'} = C_{\bbetas'}/c_{\bbetas\bbetas'}$
(by construction we are assured that $c_{\bbetas\bbetas'}\ne 0$).
It follows from
(\ref{coeffrelation}) that
${\bar\partial}\left({\hat C}_{\bbetas\bbetas_1'} -
 {\hat C}_{\bbetas\bbetas_2'}\right)=0$
for any two
$\bbeta_1',\bbeta_2' \in \partial\bb(\bbeta)$. Thus,
${\hat C}_{\bbetas\bbetas'}$ is `almost' independent of $\bbeta'$.
To define a truly $\bbeta'$-independent quantity,
let ${\hat C}_{\bbetas}$ be the average
over $\bbeta'$ of all the ${\hat C}_{\bbetas\bbetas'}$'s.
This leads to the representation
\be
   C_{\bbetas'} = {\hat C}_{\bbetas}\,c_{\bbetas\bbetas'} + C^0_{\bbetas'},
   \label{representation}
\ee
where ${\bar\partial} C^0_{\bbetas'}=0$.
Applying the various relations above to (\ref{dXbar}) we find
\bea
  &&\partial{\bar X}
  \stackrel{(\ref{dXbar},\ref{coeffrelation})}{\simeq}
  \sum_{\bbetas'\in\partial\bBbs} \left({\bar\partial}{C}_{\bbetas'}
  \right) {\bu}^{\bbetas'}
  \stackrel{(\ref{representation})}{=}
  \sum_{\bbetas\in\bBbs\atop{\scriptstyle \bbetas'\in\partial\bbs(\bbetas)}} \,
  \left({\bar\partial}{\hat C}_{\bbetas}\,c_{\bbetas\bbetas'} \right)
  {\bu}^{\bbetas'} =
  \sum_{\bbetas\in\bBbs} \left({\bar\partial}{\hat C}_{\bbetas} \right)
  \sum_{\bbetas'\in\partial\bbs(\bbetas)}\,
  c_{\bbetas\bbetas'}\,{\bu}^{\bbetas'} \nonumber\\&&
  \stackrel{(\ref{derivative})}{\simeq}
  \sum_{\bbetas\in\bBbs}\,\left({\bar\partial}{\hat C}_{\bbetas}\right)
  \partial{\bu}^{\bbetas}
  \stackrel{(\ref{balphaineq})}{\simeq}
  {\bar\partial} \sum_{\bbetas\in\bBbs}\,{\hat C}_{\bbetas}
  \partial\bu^{\bbetas}
  \stackrel{(\ref{balphaineq})}{\simeq}
  \partial{\bar\partial} \sum_{\bbetas\in\bBbs}\,{\hat C}_{\bbetas}
  \bu^{\bbetas}.
  \label{trivial2}
\eea
Equation (\ref{trivial2}) demonstrates
that the conservation law is trivial to leading order
as long as $\bab>\bo$.

  Having shown that ${\bar X}$ can be reduced to a polynomial in $\bub$
and $\bw$ through a modification (\ref{redef}), it remains to prove that
also the $\bw$-dependency can be eliminated in the same manner. We may equally
well assume the more general form ${\bar X} = {\bar X}(\bub,\bphi)$, and
show that an arbitrary dependency on $\bphi$ can be eliminated.
Assuming this form we find that
\be
  \partial{\bar X} \simeq \sum_i
  \left(\partial{\bar X}/\partial\phi_i\right) \partial\phi_i,
  \label{xbarside}
\ee
which requires that
\be
  X \simeq \sum_i C_i(\bub,\bphi) \partial\phi_i,\ \ \ \
  {\bar\partial}X \simeq \sum_i
  \left({\bar\partial}C_i\right) \partial\phi_i.
  \label{xside}
\ee
By comparing (\ref{xbarside}) and (\ref{xside}) one finds
$\partial{\bar X}/\partial\phi_i =
{\bar\partial}C_i$. Since
$\partial^2{\bar X}/\partial\phi_i\partial\phi_j=
\partial^2{\bar X}/\partial\phi_j\partial\phi_i$
this in turn implies
\be
  {\bar\partial}\left[\frac{\partial C_i}{\partial\phi_j} -
  \frac{\partial C_j}{\partial\phi_i} \right]
  \equiv
  {\bar\partial}{\cal F}_{ij} = 0.
\ee
Thus, ${\cal F}_{ij}=-{\cal F}_{ji}$ is
independent of $\bub$ and $\bphi$. We may view
$C_i$ as the electromagnetic gauge potential corresponding a constant
field strength ${\cal F}$. It is easy to verify that
$\frac{1}{2}\sum_j {\cal F}_{ij}\phi_j$ is one possible solution
for $C_i$ (provided the space is simple-connected, which we assume).
The general form must be a gauge transformation of
this potential, i.e.
\be
   C_i = \frac{1}{2} \sum_j {\cal F}_{ij} \phi_j
   + \frac{\partial D(\bub,\bphi)}{\partial\phi_i}.
\ee
Thus
\be
   X  \simeq \sum_i \frac{\partial D}{\partial\phi_i}\partial\phi_i -
   \frac{1}{2}\sum_{ij} {\cal F}_{ij} \phi_i \partial\phi_j
   \simeq \partial D - \frac{1}{2}\sum_{ij}
   {\cal F}_{ij} \phi_i\partial\phi_j.
\ee
By a modification (\ref{redef}) the $D$-term can be transformed away, and we
are left with the possibility
\bea
  && X = -\frac{1}{2}\sum_{ij} {\cal F}_{ij} \phi_i {\partial}\phi_j +
         X^0(\bub,\bphi),\nonumber\\&&
  {\bar X} = \frac{1}{2}\sum_{ij} {\cal F}_{ij} \phi_i {\bar\partial}\phi_j +
         {\bar X}^0(\bub).
\eea
However, it follows from these expressions that
\be
  {\bar\partial}X - \partial{\bar X} = \sum_{ij} {\cal F}_{ij} \phi_i v_j
  + \mbox{other terms},
  \label{notconslaw}
\ee
where the
\be
  \mbox{other terms}
  = \sum_i \frac{\partial{X}^0(\bub,\bphi)}{\partial\phi_i}\,{\bar u}_{i1} +
\sum_{ij}
  \frac{\partial X^0(\bub,\bphi)}{\partial{\bar u}_{i,j}}\,
  {\bar u}_{i+1,j}
  -\sum_{ij} \frac{\partial{\bar X}^0(\bub,\bphi)}{\partial{\bar u}_{ij}}\,
  \partial{\bar u}_{ij}
\ee
are free from terms with the combination $\phi_i v_j$.
Thus, (\ref{notconslaw}) cannot possibly
be a conservation law unless ${\cal F}_{ij}=0$.
By these arguments we have restricted ${\bar X}$ to the form
\be
   {\bar X}={\bar{\cal P}}(\bu)
   \label{genXbar}
\ee
where ${\bar{\cal P}}$ is a
homogeneous polynomial of rank $(0,{\bar h})$. From this form
it follows by the use of the equation of motion (\ref{geneqn}) that
\[
  \partial{\bar X} = \sum_{ij}
  \left(\partial{\bar X}/\partial{\bar u}_{ij}\right)\partial{\bar u}_{ij}
  = \sum_i w_i\,{\bar{\cal Q}}_i(\bub),
\]
where the ${\bar{\cal Q}}_i$'s are homogeneous polynomials of
rank $(0,{\bar h}-1)$. This in turn requires $X$ to be of the form
\be
   X = \sum_{i} w_i {\cal Q}_i(\bub),
   \label{genX}
\ee
so that
\[
  {\bar\partial}X =\sum_i w_i \left( \sum_j k_{ij} {\bar u}_{j1} +
  \sum_{j\ell} {\bar u}_{j,\ell+1} \frac{\partial}{\partial{\bar u}_{j\ell}}
  \right) {\cal Q}_i.
\]
Thus, the conservation law (\ref{conservationlaw}) is satisfied if and only
if
\be
   {\bar{\cal Q}}_i(\bub) = \left( \sum_{j} k_{ij}{\bar u}_{j1} +
   \sum_{j\ell} {\bar u}_{j,\ell+1}
   \frac{\partial}{\partial{\bar u}_{j\ell}}\right) {\cal Q}_i(\bub)
\ee
for all $i$.

As in the previous section
we finally note that the results of this section can be
extended to more general equations
\be
   \partial{\bar\partial}\bphi = - \bv,
\ee
where the forces $v_i$ may e.g.\ include polynomials in $\bphi$.
The conserved currents may still be transformed to the form
(\ref{genXbar}), (\ref{genX}), with a more general set of
$w_i$'s.
A sufficient requirement for the restriction to this form is that
\begin{enumerate}

  \item[(i)] all forces $v_i$ are expressible as linear combinations
  of derivatives $\partial w_j/\partial\phi_k$.

  \item[(i)] the set of $w_i$'s is closed under differentiations, i.e.\ that
  each $\partial w_j/\partial\phi_k$ is expressible as a linear
  combination of the $w_i$'s.

\end{enumerate}

\section{Conclusions}

In conclusion,
we have investigate the possible form of local translation invariant
conservation laws associated with the
relativistic field equations
\[
  \partial\bar\partial\phi_i=-v_i(\bphi)
\]
for a multicomponent field $\bphi$. Under the assumptions that
\begin{enumerate}

 \item[(i)]the $v_i$'s
   can be expressed as linear combinations of partial derivatives
   $\partial w_j/\partial\phi_k$ of a set of functions $w_j(\bphi)$,

 \item[(ii)]the space of functions spanned by the $w_j$'s is closed under
   partial derivations,  and

 \item[(iii)]the fields $\bphi$ take values in a simple-connected space,

\end{enumerate}
the local conservation laws can be transformed to the either form
\be
 \partial{\bar{\cal P}}=\bar\partial\sum_j w_j {\cal Q}_j,
 \label{consform}
\ee
where
$\bar{\cal P}$ and ${\cal Q}_j$ are homogeneous polynomials in the
variables
$\bar\partial\phi_i$, $\bar\partial^2\phi_i$,\ldots),
or to the parity transformed form of (\ref{consform}), obtained by
making the replacements
$\partial\equiv\partial_t+\partial_x\rightleftharpoons\bar\partial
\equiv \partial_t-\partial_x$ in all expressions.

Our results are in agreement with the conclusions of
Freedman {\em et.\ al.}\cite{Freedman}, who (among other
things) sought for ``covariant'' forms,
\[
   \partial_{\mu}{\cal J}^{\mu\alpha\cdots\omega} = 0,
\]
of the conservation laws for the 2-dimensional sine-Gordon equation.
They indeed found that the 4th order conservation law could be written
in such a covariant form, but that the additional conservation laws
were (in our sense) trivial. We have shown that this results extends
to the conservation laws of all orders, and to all field theories in the
class specified above.

\end{document}